\documentclass[aps,prl,twocolumn]{revtex4}
\usepackage[latin1]{inputenc}
\usepackage{amsmath}
\usepackage{amsfonts}
\usepackage{amssymb}
\usepackage{graphicx}
\usepackage[dvipsnames]{xcolor}
\usepackage[sort&compress]{natbib}
\usepackage{bm}

\definecolor{richlilac}{rgb}{0.71, 0.4, 0.82}

\begin{document}

	\title{Rheology and microrheology of deformable droplet suspensions}
	\author{M. Foglino, A. N. Morozov, D. Marenduzzo}
	\affiliation{SUPA, School of Physics and Astronomy, University of Edinburgh}	
	
	\begin{abstract}
	Dense suspensions of soft colloidal particles display a broad range of physical and
	rheological properties which are still far from being fully understood. To elucidate 
	the role of deformability on colloidal flow, we employ 
	computer simulations to measure the apparent viscosity of a system of droplets 
	of variable surface tension subjected to a pressure-driven flow. 
	We confirm that our suspension generically undergoes discontinuous shear thinning,
	and determine the dependence of the onset of the
	discontinuity on surface tension. 
	We find that the effective viscosity of the suspension is mainly determined by a
	capillary number. We present active microrheology simulations, where a single 
	droplet is dragged through the suspension. These also show a dynamical phase transition, 
	analogous to the one associated with discontinuous shear thinning in our interpretation. 
	Such a transition is signalled by a discontinuity in the droplet velocity versus 
	applied force.
	\end{abstract}
	
	\maketitle
	
	\section{Introduction}
	
	Emulsions are colloidal fluids with broad applications to industry and medicine, 
	including  waste treatment, food processing and pharmaceutical 
	manufacturing~\cite{Einstein_law_2,Jones}. 
	
	At a microscopic scale emulsions are concentrated colloidal suspensions made of soft 
	and deformable particles. A common example is that of blood, made of discrete and 
	deformable red blood cells (RBCs) whose microscopic properties can affect the 
	macroscopic rheological behaviour of the fluid~\cite{blood,blood2,PhysRevE.83.061924,RBC_Pagonabarraga}. 
	For instance, cells affected by malaria are commonly found to be more rigid than their unaffected counterparts. In turn, this induces a distinctive rheological behaviour of the blood~\cite{Fedosov35}. Thus, a better understanding of RBC flow under different conditions such as channel geometry, cell rigidity and shape is essential to develop and optimize new microfluidic devices, drug delivery techniques and even diagnosis protocol.
	
	A number of studies have been dedicated to investigate the physical properties of hard sphere fluids, shedding light on the physics of glass transition~\cite{Glass_transition_Cates, Glass_transition_Weeks, Glass_Pusey}, and soft glassy rheology \cite{Glassy_material_Cates_Fielding}. On the other hand, the rheological and flow properties of soft and deformable suspension \cite{AIP_droplets, AIP_droplets_2,Droplets_extensional_shear_Zinchenko, Droplets_extensional_shear_Zinchenko2, Droplets_shear_Loewenberg} remain poorly understood. 
	
	Previously~\cite{My1PRL}, we investigated the dynamics of non-coalescing deformable droplets under pressure-driven flow by means of 2D hybrid lattice Boltzmann simulations. We showed that the emulsion rheology displays a discontinuous shear thinning behaviour, which is associated with a non-equilibrium transition (or sharp crossover) between a ``hard'' and a ``soft'' regime. The former displays slow flow, caged droplet dynamics and undeformed droplets, whereas the latter is characterised by fast flow and pronounced droplet deformation. 
	
	Having identified the origin of this rich behaviour with the extent of droplet deformability, we are now interested in characterising the rheological properties as a function of the droplets surface tension. To this end, we perform hybrid lattice Boltzmann simulations of 2D emulsions in which we vary both the external pressure gradient and the softness of the particles. We show that the non-equilibrium hard-soft transition (i.e., the point of discontinuous shear thinning) is progressively shifted towards higher values of the applied pressure-difference as we increase the droplet surface tension and we find that the apparent viscosity is predominantly determined by the capillary number of the system.
	
	To further characterise the hard-soft transition, we perform simulations inspired by typical active microrheology measurements~\cite{C0CP01564D, Microrheology_Cloitre, Microrheology_Gisler, mrheol_Vadas}, where one of the droplets is selected and dragged through the suspension. This strategy enables us to study the response of the system to a localised perturbation, and we identify a discontinuous behaviour in the probe velocity with intriguing analogies to the sudden decay in viscosity displayed by the bulk.
	
	
	
	Whilst the model presented in this work may be directly recreated in the lab via suspensions of deformable and non-coalescing droplets (e.g., stabilised by a surfactant), we suggest that it could also be used to qualitatively describe the flow properties of red blood cells or other eukaryotic cells.
	
	 \section{Model and Methods}
	
	Our aim here is to investigate numerically the rheological properties of a two-dimensional suspension of non-coalescing deformable droplets subjected to a pressure-driven flow. Our simulations allow us to tune the following three key parameters which determine the flow properties of our suspension: (i) the pressure difference $\Delta P$ causing the flow, (ii) the suspension area fraction $\Phi$, defined as the ratio between the area of all droplets and the total area of the simulation domain, and (iii) the surface tension $\gamma$ which accounts for the droplets deformability.
	
	Our soft droplet suspensions are described by introducing a set of phase-fields $\phi_i$, 
	$i=1,\ldots, N$, where $N$ is the total number of droplets. The fact that we use a 
	different phase field for each droplet ensures that they are non-coalescing. On the 
	other hand, the underlying fluid flow is described by a velocity field $\mathbf{v}$.
	
	Our suspension equilibrium behaviour is described by the following free energy density:
	\begin{equation}
	\label{free_energy}
	f = \frac{\alpha}{4}\sum_i^{N}\phi_i^2(\phi_i-\phi_0)^2 + \frac{K}{2}\sum_i^{N}(\nabla\phi_i)^2  + \epsilon\sum_{i,j,i<j}\phi_i\phi_j. 
	\end{equation}
	The first term in Eq.~\eqref{free_energy} ensures stability of each of the  droplets, by
	creating two coexisting minima for $\phi = \phi_0$ and $\phi = 0$. These two minima
	represent the inside and outside area of the $i$-th droplet respectively. 
	The droplets deformability properties are determined by their surface tension 
	$\gamma =  \sqrt{8K\alpha/9}$ and can therefore be tuned, for instance, by changing the 
	value of the constant $K$. For simplicity, from now on we will refer to the latter 
	quantity $K$ as the surface tension-like parameter. The last term in the free energy 
	density introduces a soft repulsion between droplets -- its strength is measured by 
	the constant $\epsilon > 0$. This soft repulsion term prevents droplet overlap in dense
	suspensions. 
	
	The dynamics of the droplets is described by a set of Cahn-Hilliard equations for the 
	phase fields $\phi_i$:
	\begin{equation}
	\label{cahn-hilliard}
	\frac{\partial \phi_i}{\partial t} + \mathbf{\nabla} \cdot (\mathbf{v}\phi_i) = 
	M\nabla^2 \mu _i,
	\end{equation}
	where $M$ represents the mobility and $\mu_i=\partial f/\partial \phi_i-
	\partial_{\alpha}f/\partial (\partial_{\alpha}\phi)$ is the chemical potential of 
	the $i$-th droplet. This set of equation conserves the area of each droplets.
	
	The droplets dynamics is coupled to that of the underlying solvent trough the fluid 
	velocity field $\mathbf{v}$ which evolves according to the Navier-Stokes equation:
	\begin{equation}
	\label{navier-stokes}
	\rho \Big(\frac{\partial}{\partial t} + {\mathbf v}\cdot {\nabla}\Big){\mathbf v} = -{\nabla p} - \sum_i \phi_i{\nabla}\mu_i + \eta_0\nabla^2{\mathbf v},
	\end{equation}
	where $\rho$ indicates the fluid density, $p$ denotes its pressure and $\eta_0$ the 
	viscosity of the underlying solvent.  The term $\sum_i \phi_i{\nabla}\mu_i$, which can 
	be expressed as a divergence of a stress tensor, represents the internal forces due to 
	the presence of non-trivial compositional order parameters which vary spatially.
	
	We analyse the dynamics of the system by means of hybrid lattice Boltzmann simulations 
	(LB)~\cite{HybridLB_Gonnella,LB_method}, where Eq.~\eqref{navier-stokes} is solved 
	through the lattice Boltzmann algorithm while Eq.~\eqref{cahn-hilliard} is addressed 
	using a finite difference method.
	
	In our simulations, the soft droplets suspension is subjected to a constant 
	external pressure gradient acting on the underlying fluid, and causing the system 
	to flow (in a Newtonian fluid, this setup leads to Poiseuille flow with a parabolic
	velocity profile).
	In order to simulate a pressure gradient, a body force (force per unit density) 
	was included in our LB algorithm, and added to the collision operator at each lattice 
	node.  The term $\sum_i \phi_i{\nabla}\mu_i$ was also added as a body force, as this 
	procedure limits the spurious LB velocities in equilibrium~\cite{SpuriousVelocities_Wagner}. 
	Our simulations are run on a $96 \times 96$ lattice, with periodic boundary conditions 
	along the $x$ axis , and boundary walls along the $y$ axis. Neutral wetting conditions
	were used, as they create a layer of boundary droplets which slow down the motion
	of those which come into contact with them. These were enforced by requiring that, 
	for each $i = 1,..., N$,

	\begin{equation}
	\frac{\partial \mu_i}{\partial z} = 0, \qquad
	\frac{\partial \nabla ^2 \phi_i}{\partial z} = 0.
	\label{BC}
	\end{equation}

	In Eq.\eqref{BC}, the first condition ensures density conservation (no flux boundary), 
	while the second one determines the wetting to be neutral, i.e. the droplets at 
	the boundary form an angle of $90^\circ$ with the wall surface. 
	
	The parameters used in our simulations are the following. We fix the droplet radius to 
	$r = 8$ and the mobility to $M = 0.1$, while the free energy parameters are 
	$\alpha = 0.07$, $\epsilon=0.05$, and $K$ varies within the range $[0.02-0.25]$. 
	The viscosity of the underlying fluid is $\eta_0 = \frac{5}{3}$ (for simplicity this 
	is also the viscosity of the fluid inside the droplets), and the value of applied 
	pressure difference varies within the range $[10^{-5} - 10^{-4}]$. While the trends 
	that we show in our results section are generic and do not rely on a particular choice 
	of physical unit, the parameters listed above can be mapped onto a physical system with 
	droplets of size (diameter)  equal to $100\mu m$, in a background fluid with viscosity 
	$10^{-2} Pa s$ (in the absence of droplets), and where the surface tension 
	$\gamma = \sqrt{(8K\alpha)/9} \sim 0.09$ corresponds to $1 mN/m$. With this mapping, 
	a velocity of $10^{-3}$ in simulation units corresponds to $1 mm/s$. The Reynolds number 
	ranges from $\sim 0.04$ ($\Delta P = 5 \times 10^{-5}$, resulting in a fluid maximum 
	velocity $v_{max} \sim 0.03$) to $\sim 8$ (for $\Delta P = 10^{-4}, 
	v_{max} \sim 0.07$ ). These values are always small enough that  no inertial 
	instabilities are observed. The conventional Capillary number $Ca = \frac{\eta_0v}
	{\gamma}$ ranges from $\sim 0.06$ (for $\Delta P = 5 \times 10^{-5}$) to $\sim 1.23$ 
	(for $\Delta P = 10 ^{-4}$). 
	
	\section{Results}
	
	\subsection{Pressure-driven flow}
	
	\begin{figure}[h!]
		\centering
		\hspace{-10pt}\includegraphics[width=.50\textwidth]{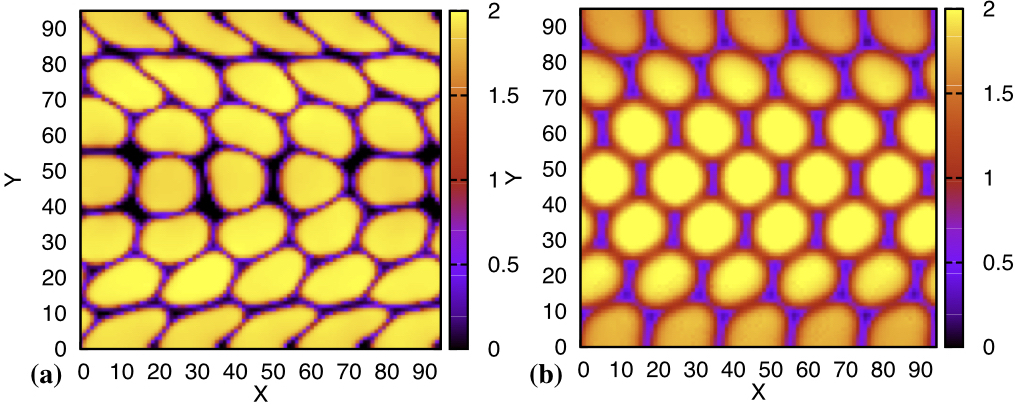}
		\setcounter{figure}{0} 
		\caption{(\small \textbf{(a)} Snapshot of the simulation of a suspension of $\Phi = 76.3 \%$, $K = 0.02$ subjected to a pressure difference $\Delta P = 7 \times 10^{-5}$.\textbf{(b)} Snapshot of the simulation of a suspension of $\Phi = 76.3 \%$, $K = 0.18$ subjected to a pressure difference $\Delta P = 7 \times 10^{-5}$. It can be seen that lower values of $K$ lead to a softer foam, with more pronounced droplet deformations.}
		\label{Fig_1}
	\end{figure}
	
	In this work we aim to understand and characterise the role of droplet deformability 
	in determining the flowing properties of the overall suspension. 
	To this end, we perform two-dimensional simulations of a suspension of deformable 
	droplets in a liquid medium, subjected to a pressure-driven flow, where we 
	systematically vary the values of droplet surface tension $K$ and applied pressure 
	difference $\Delta P$. 
	A first feel for the extent to which surface tension can affect the 
	behaviour of the system comes from inspection of  Fig.~\ref{Fig_1}, 
	where the simulation snapshots for two different values of $K$ 
	display a remarkable qualitative difference.
 	
	To quantitatively characterise the rheological behaviour of our suspension, we measure
	the throughput flow  in our simulations, $Q = \int dy v_x(y)$, and use it to define an
	effective viscosity $\eta_{eff} = \frac{\eta}{\eta_0}$,  where $\eta_0$ is the 
	viscosity of the underlying fluid. This effective viscosity is defined as the ratio 
	between the throughput flow in a Newtonian fluid with viscosity $\eta_0$ and $Q$.
	As shown in Fig.~\ref{Fig_2}\textbf{(a)} and \textbf{(b)}, $\eta_{\rm eff}$ 
	undergoes a discontinuous shear thinning behaviour both as a function of applied 
	pressure difference $\Delta P$ and surface tension-like parameter $K$. 
	Additionally, the various viscosity curves in Fig.~\ref{Fig_2}\textbf{(a)}, referring 
	to different values of the surface tension-like parameter $K$, show that the
	value of the applied $\Delta P$ corresponding to the discontinuous shear thinning
	progressively shifts towards higher values as $K$ is increased.
	
	This discontinuous shear thinning behaviour, as discussed in our previous 
	work~\cite{My1PRL}, is a result of the system transition between a ``hard'' and a 
	``soft'' phase. While the former is characterised by almost spherical droplets which 
	tend to resist deformation and flow very slowly, the latter is associated with droplets 
	that are more susceptible to deformation and therefore flow more easily. Such an 
	interpretation is in agreement with the observed shift of the pressure difference
	corresponding to the viscosity jump, $\Delta P_{jump}$, as a function of $K$ shown the 
	inset of Fig.~\ref{Fig_2}\textbf{(a)}. Higher values of $K$ cause the suspension to be 
	overall less deformable, and we therefore expect the transition towards the soft phase 
	to occur for progressively larger values of applied pressure difference, as  
	larger forces are needed to create droplet deformations. 
	We can use a similar argument to explain the behaviour observed in 
	Fig.~\ref{Fig_2}\textbf{(b)} where the suspension viscosity is now plotted as a 
	function of $K$ for different fixed value of applied pressure difference $\Delta P$. 
	In this case, increasing $K$ takes the system from the soft to the hard phase, 
	so that the viscosity sharply increases. The jump in the suspension effective 
	viscosity shifts towards higher $K$ values as we increase the applied $\Delta P$, 
	as the latter parameter favours the soft phase. Additionally, increasing $\Delta P$
	leads to overall smaller effective viscosities overall, as the system is shear 
	thinning. 
	
	\begin{figure}[h!]
		\centering
		\hspace{-15pt}\includegraphics[width=0.51\textwidth]{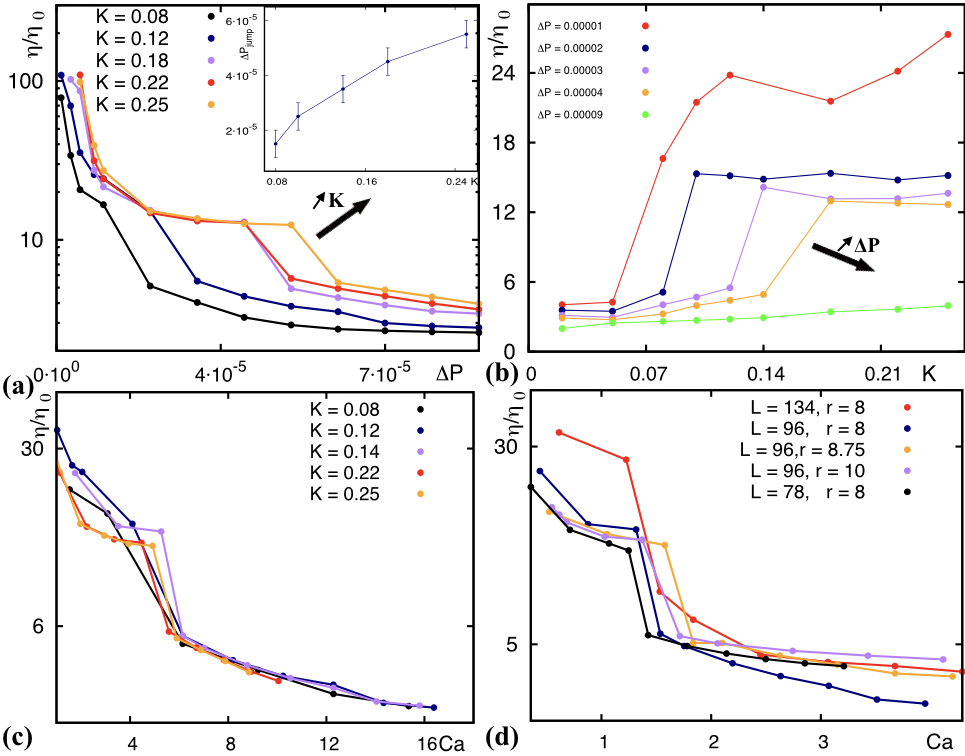}
		\setcounter{figure}{1} 
		\caption{\textbf{(a)}Plot of the suspension effective viscosity $\eta_{eff}$ as a function of the applied pressure difference for a system of area fraction $\Phi = 65.4\%$ and different fixed values of the surface tension-like parameter $K$. \textbf{Inset} Plot of the values of $\Delta P$ corresponding the the viscosity drop as a function of $K$.\textbf{(b)}Plot of $\eta_{eff}$ as a function of the surface tension-like parameter $K$ for a system of area fraction $\Phi = 65.4\%$ and different fixed values of the applied pressure difference $\Delta P$. \textbf{(c)}Plot of the suspension effective viscosity as a function of the system Capillary number $Ca$ for a system of area fraction $\Phi = 65.4\%$ and different fixed values of $K$. Comparing \textbf{(c)} with \textbf{(a)}, we can notice that the various viscosity curves collapse onto a single one if they are plotted as a function of $Ca$ instead of $\Delta P$.  \textbf{(d)} Plot of the suspension effective viscosity   as a function of the Capillary number $Ca$ for a system of area fraction $\Phi = 64.5\%$ and different fixed values of the droplet radius $r$ and system size $L$. The scaling is compatible with $r^{2}L$.}
		\label{Fig_2}
	\end{figure}
	
	We found empirically that within the parameter range the shape of the viscosity curves is predominantly determined by the following capillary number, 
	\begin{equation}
	Ca = \frac{\Delta P}{K} Lr^2
	\label{ca}
	\end{equation}
	where $r$ is the droplet radius and $L$ is the system size. Of all functional forms of the type $\frac{\Delta P}{K} L^nr^{(3-n)}$ with $n$ integer, this is the combination which leads to the best collapse~\footnote{While simulations in the main text have been performed with $\alpha=0.07$, we have also performed simulations at three other values of $\alpha$. Whilst this parameter does not enter the capillary number, it can still affect the viscosity curves as, for instance, increasing it leads to a decrease in the interfacial thickness of the droplets, hence to a decrease in the effective area fraction and, therefore, in the viscosity.}.
	Note that, while this ratio still measures the relative contribution of viscous and interfacial forces, the latter are now proportional to $K$, the term which enters the free energy density, rather than $\sqrt{K}$ as in the conventional definition of the capillary number (given above).
	The fitness of such a capillary number as a key parameter describing the 
	rheological properties of our suspension is demonstrated by 
	Fig.~\ref{Fig_2}\textbf{(c)} and \textbf{(d)}, where the effective viscosity curves referring to 
	different values of fixed droplet surface tension are plotted as a function of $Ca$. 
	As is apparent from a comparison between Fig.~\ref{Fig_2}\textbf{(a)} and \textbf{(c)}, 
	the viscosity curves for different values of $K$ all approximately collapse 
	on the same master curve as we plot them against the Capillary number (instead of 
	the simple applied pressure difference $\Delta P$). This is in line with
	the results obtained in~\cite{RBC_Pagonabarraga} for simulations of red blood cell
	rheology, within a model exhibiting continuous shear thinning.
	Moreover, Fig.~\ref{Fig_2}\textbf{(d)} shows that the suspension viscosity curves display a scaling with $Ca$ proportional to $r^2L$.
	We therefore conclude that the Capillary number is the appropriate control parameter 
	for describing the elastic and flow properties of our suspension.

	\begin{figure}[h!]
		\centering
		\hspace{-10pt}\includegraphics[width=0.5\textwidth]{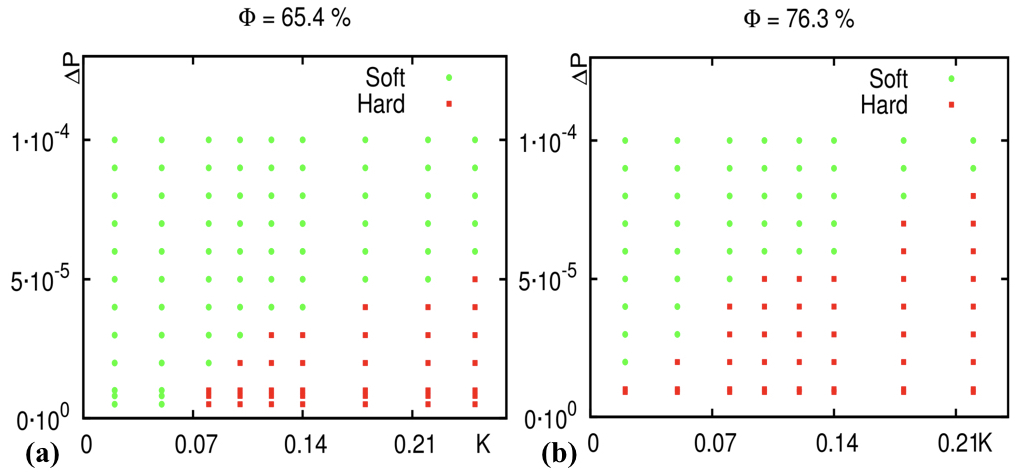}
		\setcounter{figure}{2} 
		\caption{\textbf{(a)}Phase diagram for a suspension of area fraction $\Phi = 65.4\%$ obtained through simulations performed over different values of surface tension-like parameter $K$ and applied pressure difference $\Delta P$. The green area represents the soft phase of our suspension, while the red one corresponds to the hard regime.\textbf{(b)}Phase diagram as in \textbf{(a)} for a system of area fraction $\Phi = 76.3\%$.}
		\label{Fig_3}
	\end{figure}
	
	\begin{figure*}
		\centering
		\includegraphics[scale = 0.50]{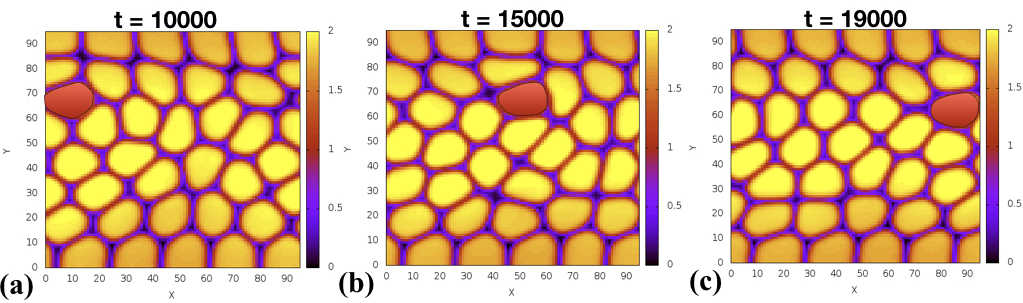}
		\setcounter{figure}{3} 
		\caption{Snapshots from a simulation of active micro-rheology. The red droplet (the colour of the probe droplet has been manually modified in order to mark it, the actual value of $\phi$ is $2$ like all the other droplets) is the probe which moves due to the applied body-force $f = 6\cdot10^{-5}$. Snapshots \textbf{(a)}, \textbf{(b)} and  \textbf{(c)} are taken from the same simulation and refer to different timesteps (namely $t = 10000, 15000$ and $t = 19000$). As we can notice, the probe droplet travels across the entire simulation box within $9000$ timesteps. }
		\label{Fig_micro_2}
	\end{figure*}
	
	Having performed simulations for a variety of different parameters, we are now able to 
	draw a phase diagram describing the transition between the soft and hard phase of our 
	system. As previously explained, the transition between these two phases is marked by a 
	sudden drop in the suspension effective viscosity for a certain threshold of applied 
	pressure difference. We therefore label as ``soft'' all the cases where the value of 
	viscosity is ``low'' -- i.e., corresponding to the right branch of the viscosity
	curve, following the downward jump corresponding to discontinuous shear thinning 
	(see Fig.~\ref{Fig_2}\textbf{(a)}). On the other hand, all the cases belonging to 
	the left branch of the viscosity curve -- i.e., to higher values of $\eta_{eff}$ -- 
	are labelled as ``hard''. In Fig.~\ref{Fig_3} we plot two phase diagrams corresponding 
	to different values of the suspension area fraction $\Phi$, with increasing values 
	of the surface tension-like parameter on the x axis, and of the applied pressure 
	difference on the y axis. The green and red area corresponds to the soft and hard 
	regime, respectively.
	As expected intuitively, as we move along the $x-$axis -- i.e., increasing $K$ and 
	keeping fixed $\Delta P$ -- our system switches from the soft towards the hard phase. 
	On the other hand, moving along the y axis -- i.e., increasing $\Delta P$ keeping 
	the value of $K$ fixed -- droplet deformation increases and the suspension undergoes
	a hard-to-soft transition. 
	
	The comparison between Fig.~\ref{Fig_3}\textbf{(a)} and \textbf{(b)} additionally 
	suggests that the area fraction $\Phi$ affects the overall stability of the soft 
	and hard phases. Thus, for the lower $\Phi$ the soft phase is predominant
	(green area in Fig.~\ref{Fig_3}\textbf{(a)}), whereas for the larger $\Phi$
	the range of stability of the hard phase increases in size (red area in 
	Fig.~\ref{Fig_3}\textbf{(b)}). This is because increasing $\Phi$ promotes
	jamming and a solid-like (hence hard) behaviour. Note that this is true despite
	the fact that at large $\Phi$ particles are frequently non-circular: what is important
	for the hard phase to be stable is that subsequent time-dependent deformations 
	from this ground state are unlikely.
	
	\subsection{Micro-rheology}
	
	We now discuss a different set of simulations, this time
	modelling a microrheology experiment.
	Such experiments analyse the dynamics of a single probe particle, or droplet, which is
	immersed in a suspension. This setup makes it possible to extract some rheological 
	properties of the overall suspension avoiding the limitation of a bulk rheology 
	measurement (such as the required big sample size).
	
	Here we focus on the case of {\it active} microrheology, where one of the droplets is
	selected and dragged through the suspension -- we call this the `probe' droplet. 
	To simulate this case, we apply a spatially dependent body-force to the system --
	along the horizontal direction in Fig.~\ref{Fig_micro_2}. This body force is modulated
	by the phase field of the probe droplet. This procedure allow us to force motion of
	the probe droplet through the suspension, and then monitor how its dynamics
	is affected by $\Phi$, $K$ and the magnitude of the force
	(more precisely, its maximum over space) which we refer to as $f$.
	While in our previous analysis, the pivotal quantity used to characterise the suspension 
	rheological properties was its effective viscosity, we now study how the (steady
	state) probe velocity changes with parameters. 	
	Remarkably, we find that discontinuous shear thinning leaves a detectable signature
	in microrheology measurements as well. Thus, as shown in Fig.~\ref{Fig_micro}, we observe
	a discontinuous behaviour in the curve portraying the probe velocity curve as a 
	function of the applied body-force. In analogy with the discontinuity in the effective 
	viscosity, such jump in the probe velocity signals a transition from a hard,
	solid-like, regime, to a soft, liquid-like, one. 
	For low values of $f$ the probe struggles to move steadily, and is in practice 
	trapped by its neighbouring droplets. On the other hand, there is a critical 
	threshold of $f$ after which the probe droplet is able to move much more easily 
	across the suspension, as it can deform more readily to squeeze past the 
	other droplets in the suspension.
	
	\begin{figure*}
		\centering
		\includegraphics[scale = 0.5]{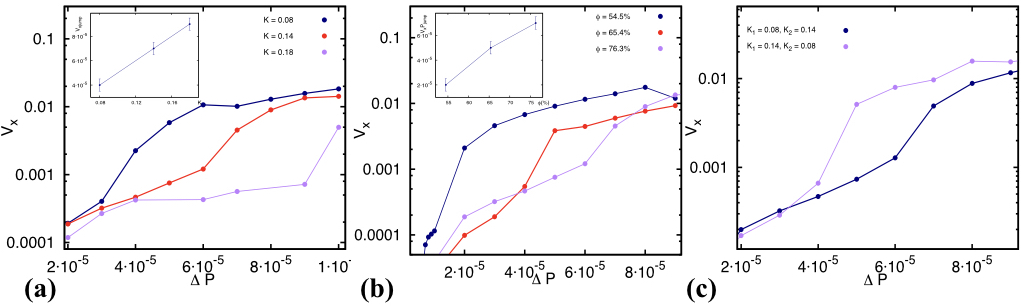}
		\setcounter{figure}{4} 
		\caption{\textbf{(a)}Probe droplet velocity curves as a function of the applied forcing for a suspension area fraction of $\Phi = 76.3\%$. The three curves refer to different values of the surface tension-like parameter, namely $K = 0.08$, $K = 0.14$ and $K = 0.18$. As we can notice, the jump in the probe droplet velocity is progressively shifted towards higher values of the forcing as $K$ is increased. \textbf{Inset} Plot of the threshold value of the forcing, associted with the velocity discontinuity, as a function of $K$. \textbf{(b)} Probe droplet velocity curves as a function of the applied forcing for a suspension of surface tension $K = 0.14$. The three curves refer to different values of the suspension area fraction, namely $\phi = 54.5\%$, $\phi = 65.4\%$ and $\phi = 76.3\%$. \textbf{Inset} Plot of the threshold value of the forcing which is associated with the velocity discontinuity as a function of the suspension area fraction. \textbf{(c)} Probe droplet velocity curves as a function of the applied forcing for a suspension of area fraction $\phi = 76.3\%$ for different values of the surface tension-like parameter. In these simulations, the probe is characterised by a different $K$ ($K_1$) with respect to the rest of the rest of the droplets ($K_2$). The purple curve refers to the case where the we simulate a ``hard'' probe moving across a ``soft'' suspension, while the blue one correspond to the opposite case.}
		\label{Fig_micro}
	\end{figure*}
	
	As discussed in the previous section, in a mascroscopic flow setup, the location of 
	the hard-to-soft transition can be tuned by the value of the droplet surface tension-like
	parameter $K$, as demonstrated by the progressive shift of the viscosity discontinuity 
	towards higher values of pressure difference as $K$ is increased.
	Is a similar effect seen in microrheology? To address this question, we 
	analysed how the location of the discontinuity in the probe droplet velocity
	is affected by changes in $K$. The results of these simulations, presented in 
	Fig.~\ref{Fig_micro}, show that the discontinuity occurs for progressively 
	larger values of the applied forcing as we increase the value of $K$. 
	In agreement with our previous macroscopic flow results, the transition from 
	the hard to the soft phase becomes more ``difficult'' for suspensions with stiffer
	droplets: i.e., in order to reach the soft phase a higher value of $\Delta P$ is 
	needed. In the inset of Figs.\ref{Fig_micro}\textbf{(a)} and \textbf{(b)} we show 
	the value of the forcing correspondent to the jump in the probe velocity as 
	a function of the surface tension and area fraction, respectively.  As can be seen, 
	this value increases near-linearly as a function of both parameters, and 
	increasing either $K$ or $\Phi$ favours the hard phase. 
	
	Finally, we performed simulations where we set different values for the surface tension
	of the probe droplet and the rest of the droplets in the suspension. The rationale
	for doing so is to investigate whether the physical features of the probe affect our
	measurements, and in particular our estimate for the threshold forcing needed to
	trigger the transition to the soft phase. For an ideal active 
	microrheological experiments, the terminal velocity of the probe should be
	determined by the Stokes law, hence only depend on probe size and medium effective
	viscosity (where the latter should ideally be probe-independent).
	
	We considered two cases: (i) one in which case a soft probe is dragged across a 
	suspension of more rigid droplets, and (ii) the opposite scenario where
	a hard probe moves within a more deformable environment. As shown in 
	Fig.~\ref{Fig_micro}\textbf{(c)}, these two complementary cases yield a 
	different threshold for the forcing which triggers the hard-to-soft transition. 
	When a soft probe ($K_1 = 0.08$) is immersed in a more rigid suspension 
	($K_2 = 0.14$), the transition occurs for a value of $f$ which is indistinguishable 
	from that obtained previously for the case of homogeneous 
	$K = 0.14$ ($f = 7\cdot 10^{-5}$). When considering a 
	rigid probe ($K_1 = 0.14$) moving across a softer suspension ($K_2 = 0.08$), 
	the threshold of the forcing shifts towards a lower value 
	($f = 5\cdot 10^{-5}$), again suggesting a similar behaviour to the overall 
	soft suspension case with homogeneous surface tension-like parameter, now
	$K = 0.08$. Our results therefore suggest that, within our microrheology
	set-up, the details of the probe do not significantly affect the measurements,
	and in particular the estimate of the hard-to-soft transition point --
	the droplet being dragged therefore behave as an ideal microrheological probe in
	this respect. 
	
	
	\section{Conclusions}
	
	In summary, we presented here two dimensional lattice Boltzmann simulations of a system 
	of non-coalescing, deformable droplet suspension where the role of droplet surface 
	tension and deformability are investigated. Applying a pressure-driven macroscopic flow, 
	we observe that the rheology of our system entails a discontinuous shear thinning 
	behaviour, which is associated with a non-equilibrium phase transition (or very sharp 
	crossover) between a hard phase, which flows slowly, and a soft phase, where more 
	frequent droplet deformations lead to a faster flow. In order to test the effect of
	droplet deformability more systematically, we performed a set of simulations where we 
	study the behaviour of the system (effective viscosity) as a function of both the
	applied pressure difference as well as the droplet surface tension. Our results show 
	that the observed hard-soft transition progressively shifts towards higher values 
	of applied pressure difference as we increase the droplet surface tension. 
	In other words as shown in Fig.\ref{Fig_2}\textbf{(a)} and Fig.~\ref{Fig_3}, a more 
	rigid system (high values of $K$) requires a higher threshold value of the applied 
	pressure difference to trigger its transition towards the soft phase. 
	
	Notably, we found that the rheological properties of our suspension are mainly determined
	by its capillary number $Ca$, defined for our system as in \eqref{ca}. This parameter, 
	which captures the interplay between the surface tension and external flow, can be used
	to approximately collapse all the flow (viscosity) curves onto a single master curve, 
	as shown in Fig.~\ref{Fig_2}\textbf{(c)}. This suggests that $Ca$ is the main parameter
	determining the macroscopic flow of our suspensions. 
	
	Finally, we performed active microrheology simulations, where a single probe droplet is 
	dragged across the overall suspension. By varying the value of the body force applied to
	the probe, we observed a discontinuous behaviour in its steady state (terminal) velocity.
	This discontinuity is the analogue of the sharp viscosity drop observed in our macroscopic
	flow simulations, and corresponds to a sharp decrease in the effective friction felt
	by the probe droplet. We interpret this phenomenon as another signature of the
	nonequilibrium hard-to-soft transition that our suspension undergoes.
	
	We thank Davide Michieletto for numerous useful discussions, and we acknowledge  ERC (COLLDENSE Network) for funding.

	\newpage
	
	\bibliographystyle{unsrt}
	\bibliography{bibliography}  
	\renewcommand{\thefigure}{S\arabic{figure}}

\end{document}